\documentclass{ws-ijmpa}

\begin{document}

\markboth{Hai-Yang Cheng} {Charmless 3-body $B$ decays}

%%%%%%%%%%%%%%%%%%%%% Publisher's Area please ignore %%%%%%%%%%%%%%%
%
\catchline{}{}{}{}{}
%
%%%%%%%%%%%%%%%%%%%%%%%%%%%%%%%%%%%%%%%%%%%%%%%%%%%%%%%%%%%%%%%%%%%%

\title{Charmless 3-body $B$ Decays: Resonant and Nonresonant Contributions }

\author{Hai-Yang Cheng}

\address{Institute of Physics, Academia Sinica\\
Taipei, Taiwan 115, ROC\\}

\maketitle

\begin{abstract}
Charmless 3-body decays of $B$ mesons are studied using a simple
model based on the framework of the factorization approach. We have
identified a large source of the nonresonant signal in the matrix
elements of scalar densities, e.g. $\langle K\overline K|\bar
ss|0\rangle$. This explains the dominance of the nonresonant
background in $B\to KKK$ decays, the sizable nonresonant fraction of
order $(35\sim 40)\%$ in $K^-\pi^+\pi^-$ and $\overline
K^0\pi^+\pi^-$ modes and the smallness of nonresonant rates in $B\to
\pi\pi\pi$ decays. We have computed the resonant and nonresonant
contributions to charmless 3-body decays and determined the rates
for the quasi-two-body decays $B\to VP$ and $B\to SP$.
Time-dependent $CP$ asymmetries $\sin2\beta_{\rm eff}$ and $A_{CP}$
in $K^+K^-K_S,K_SK_SK_S,K_S\pi^+\pi^-$ and $K_S\pi^0\pi^0$ modes are
estimated.

%\keywords{Keyword1; keyword2; keyword3.}
\end{abstract}

%\ccode{PACS numbers: 11.25.Hf, 123.1K}

\section{Introduction}

Recently many three-body $B$ decay modes have been observed with
branching ratios of order $10^{-5}$. The Dalitz plot analysis of
3-body $B$ decays provides a nice methodology for extracting
information on the unitarity triangle in the standard model. The
three-body meson decays are generally dominated by intermediate
vector and scalar resonances, namely, they proceed via
quasi-two-body decays containing a resonance state and a
pseudoscalar meson. Indeed, most of the quasi-two $B$ decays are
extracted from the analysis of three-body $B$ decays using the
Dalitz plot technique.

\begin{table}[h]
\tbl{Branching ratios (in units of $10^{-6}$) of nonresonant
contributions to various charmless three-body decays of $B$
mesons~$^{1}$. The nonresonant fractions (in \%) are shown in
parentheses.}
{\begin{tabular}{@{}lcc@{}} \toprule Decay & BaBar & Belle  \\
\colrule
$B^-\to\pi^+\pi^-\pi^-$\hphantom{00} & \hphantom{0}$2.3\pm0.9\pm0.5~(13.6\pm6.1)$ & \hphantom{0}  \\
$B^-\to K^-\pi^+\pi^-$\hphantom{00} & \hphantom{0}$2.87\pm0.65\pm0.43^{+0.63}_{-0.25}~(4.5\pm1.5)$ &
\hphantom{0}$16.9\pm1.3\pm1.3^{+1.1}_{-0.9}~(34.0\pm2.9)$ \\
$\overline B^0\to\overline K^0\pi^+\pi^-$\hphantom{0} &  & $19.9\pm2.5\pm1.6^{+0.7}_{-1.2}~(41.9\pm5.5)$ \\
$B^-\to K^+K^-K^-$ & $50\pm6\pm4~(141\pm18)$  & $24.0\pm1.5\pm1.5~(74.8\pm3.6)$ \\
$\overline B^0\to K^+K^-\overline K^0$ & $26.7\pm4.6~(112\pm15)$ &  \\
\botrule
\end{tabular} \label{tab:BRexpt}}
\end{table}

It is known that the nonresonant signal in charm decays is small,
less than 10\% \cite{PDG}. In the past few years, some of the
charmless $B$ to 3-body decay modes have been measured at $B$
factories and studied using the Dalitz plot analysis. We see from
Table \ref{tab:BRexpt} that the nonresonant fraction is about 90\%
in $B\to KKK$ decays, $35\sim 40\%$ measured by Belle and 5\% by
BaBar in $B\to K\pi\pi$ decays and 14\% in the $B\to\pi\pi\pi$
decay. Hence, the nonresonant 3-body decays could play an essential
role in $B$ decays. While this is a surprise in view of the rather
small nonresonant contributions in 3-body charm decays, it is not
entirely unexpected because the energy release scale in weak $B$
decays is of order 5 GeV, whereas the major resonances lie in the
energy region of 0.77 to 1.6 GeV. Consequently, it is likely that
3-body $B$ decays may receive sizable nonresonant contributions.

\section{Resonant and Nonresonant contributions}
We take the decay $\overline B^0\to K^+K^-\overline K^0$ as an
illustration. Under the factorization approach, its decay amplitude
consists of three distinct factorizable terms: (i) the transition
process induced by $b\to s$ penguins, $\langle \overline B^0\to
K^+\overline K^0\rangle\times \langle 0\to K^-\rangle$,  (ii) the
current-induced process through the tree $b\to u$ transition,
$\langle \overline B^0\to \overline K^0\rangle\times \langle 0\to
K^+K^-\rangle$,  and (iii) the annihilation process $\langle
\overline B^0\to 0\rangle\times \langle 0\to K^+K^-\overline
K^0\rangle$, where $\langle A\to B\rangle$ denotes a $A\to B$
transition matrix element.

\subsection{Nonresonant background}
For the transition process, the general expression of the
nonresonant contribution has the form
 \begin{eqnarray} \label{eq:AHMChPT}
 && \langle K^-(p_3)|(\bar s
 u)_{V-A}|0\rangle \langle\overline K {}^0 (p_1) K^+(p_2)|(\bar u b)_{V-A}|\overline B {}^0\rangle^{NR} \nonumber\\
 &=& -\frac{f_K}{2}\left[2 m_3^2 r+(m_B^2-s_{12}-m_3^2) \omega_+
 +(s_{23}-s_{13}) \omega_-\right],
 \end{eqnarray}
where $(\bar q_1q_2)_{V-A}\equiv \bar
q_1\gamma_\mu(1-\gamma_5)q_2$. In principle, one can apply heavy
meson chiral perturbation theory (HMChPT) to evaluate the form
factors $r,~\omega_+$ and $\omega_-$ (for previous studies, see \cite{Fajfer}). However, this will lead to
too large decay rates in disagreement with experiment
\cite{Cheng:2002qu}. A direct calculation indicates that the
branching ratio of $\overline B^0\to K^+K^-\overline K^0$ arising
from the transition process alone is already at the level of
$77\times 10^{-6}$ which exceeds the measured total branching ratio
\cite{HFAG} of $25\times 10^{-6}$. The issue has to do with the
applicability of HMChPT. In order to apply this approach, two of
the final-state pseudoscalars ($K^+$ and $\overline K^0$ in this
example) have to be soft. The momentum of the soft pseudoscalar
should be smaller than the chiral symmetry breaking scale
$\Lambda_\chi$ of order $0.83-1.0$ GeV. For 3-body charmless $B$
decays, the available phase space where chiral perturbation theory
is applicable is only a small fraction of the whole Dalitz plot.
Therefore, it is not justified to apply chiral and heavy quark
symmetries to a certain kinematic region and then generalize it to
the region beyond its validity. If the soft meson result is assumed
to be the same in the whole Dalitz plot, the decay rate will be
greatly overestimated.

Recently we propose to parametrize the $b\to u$ trasnition-induced
nonresonant amplitude given by Eq. (\ref{eq:AHMChPT}) as
\cite{CCS3body}
 \begin{eqnarray} \label{eq:ADalitz}
  A_{\rm NR}=A_{\rm NR}^{\rm
  HMChPT}\,e^{-\alpha_{_{\rm NR}}
p_B\cdot(p_1+p_2)}e^{i\phi_{12}},
 \end{eqnarray}
so that the HMChPT results are recovered in the chiral limit
$p_1,~p_2\to 0$. That is, the nonresonant amplitude in the soft
meson region is described by HMChPT, but its energy dependence
beyond the chiral limit is governed by the exponential term
$e^{-\alpha_{_{\rm NR}} p_B\cdot(p_1+p_2)}$. The unknown parameter
$\alpha_{_{\rm NR}}$ can be determined from the data of the
tree-dominated decay $B^-\to\pi^+\pi^-\pi^-$. Experimentally, a
phenomenological parametrization of the non-resonant $B\to KKK$
amplitudes described by
 \begin{eqnarray} \label{eq:ANR}
A_{\rm NR}=(c_{12}e^{i\phi_{12}}e^{-\alpha
s_{12}^2}+c_{13}e^{i\phi_{13}}e^{-\alpha
s_{13}^2}+c_{23}e^{i\phi_{23}}e^{-\alpha s_{23}^2})(1+b_{\rm
NR}e^{i(\beta+\delta_{\rm NR})})
 \end{eqnarray}
is adopted by both BaBar and Belle.

In addition to the $b\to u$ tree transition, we need to consider the
nonresonant contributions to the $b\to s$ penguin amplitude
 \begin{eqnarray}
 A_1 &=& \langle \overline K {}^0|(\bar s b)_{V-A}|\overline B {}^0\rangle
  \langle K^+ K^-|(\bar u u)_{V-A}|0\rangle, \nonumber \\
 A_2 &=& \langle \overline K {}^0|\bar s b|\overline B {}^0\rangle
       \langle K^+ K^-|\bar s s|0\rangle.
 \end{eqnarray}
The 2-kaon creation matrix elements can be expressed in terms of
time-like kaon current form factors as
 \begin{eqnarray}\label{eq:KKweakff}
 \langle K^+(p_{K^+}) K^-(p_{K^-})|\bar q\gamma_\mu q|0\rangle
 &=& (p_{K^+}-p_{K^-})_\mu F^{K^+K^-}_q,
 \nonumber  \\
 \langle K^0(p_{K^0}) \overline K^0(p_{\bar K^0})|\bar q\gamma_\mu
 q|0\rangle
 &=& (p_{K^0}-p_{\bar K^0})_\mu F^{K^0\bar K^0}_q.
 \end{eqnarray}
The weak vector form factors $F^{K^+K^-}_q$ and $F^{K^0\bar K^0}_q$
can be related to the kaon electromagnetic (e.m.) form factors
$F^{K^+K^-}_{em}$ and $F^{K^0\bar K^0}_{em}$ for the charged and
neutral kaons, respectively. Phenomenologically, the e.m. form
factors receive resonant and nonresonant contributions
 \begin{eqnarray} \label{eq:KKemff}
 F^{K^+K^-}_{em}= F_\rho+F_\omega+F_\phi+F_{NR}, \qquad
 F^{K^0\bar K^0}_{em}= -F_\rho+F_\omega+F_\phi+F_{NR}'.
 \end{eqnarray}
The resonant and nonresonant terms in Eq. (\ref{eq:KKemff}) can be
determined from a fit to the kaon e.m. data. The non-resonant
contribution to the matrix element $\langle K^+ K^-|\bar s
s|0\rangle$ is given by
 \begin{eqnarray} \label{eq:KKssme}
 \langle K^+(p_2) K^-(p_3)|\bar s s|0\rangle^{NR}
 \equiv f_s^{K^+K^-}(s_{23})=\frac{v}{3}(3 F_{NR}+2F'_{NR})+\sigma_{_{\rm NR}}
 e^{-\alpha\,s_{23}},
 \end{eqnarray}
The nonresonant $\sigma_{_{\rm NR}}$ term is introduced for the
following reason. Although the nonresonant contributions to
$f_s^{KK}$ and $F_s^{KK}$ are related through the equation of
motion, the resonant ones are different and not related {\it a
priori}. As stressed in \cite{CCSKKK}, to apply the equation of
motion, the form factors should be away from the resonant region. In
the presence of the resonances, we thus need to introduce a
nonresonant $\sigma_{_{\rm NR}}$ term which can be constrained by
the measured $\overline B^0\to K_SK_SK_S$ rate and the $K^+K^-$ mass
spectrum \cite{CCS3body}.

\subsection{Resonant contributions}
Vector meson and scalar resonances contribute to the two-body matrix
elements $\langle P_1P_2|V_\mu|0\rangle$ and $\langle
P_1P_2|S|0\rangle$, respectively. They can also contribute to the
three-body matrix element $\langle P_1P_2|V_\mu-A_\mu|B\rangle$.
Resonant effects are described in terms of the usual Breit-Wigner
formalism. More precisely,
 \begin{eqnarray}
 \langle K^+K^-|\bar q\gamma_\mu q|0\rangle^R &=& \sum_i\langle K^+K^-|V_i\rangle
{1\over m_{V_i}^2-s-im_{V_i}\Gamma_{V_i}}\langle V_i|\bar
q\gamma_\mu
q|0\rangle, \nonumber \\
 \langle K^+K^-|\bar ss|0\rangle^R &=& \sum_i\langle K^+K^-|S_i\rangle
{1\over m_{S_i}^2-s-im_{S_i}\Gamma_{S_i}}\langle S_i|\bar
ss|0\rangle,
 \end{eqnarray}
where $V_i=\phi,\rho,\omega,\cdots$ and
$S_i=f_0(980),f_0(1370),f_0(1500),\cdots$.
In this manner we are able to figure out the relevant resonances
which contribute to the 3-body decays of interest and compute the
rates of $B\to VP$ and $B\to SP$.

\section{Penguin-dominated $B\to KKK$ and $B\to K\pi\pi$ decays}

As mentioned in the previous section, we employ the decays
$\overline B^0\to K^+K^-\overline K^0$ and $K_SK_SK_S$ to fix the
nonresonant parameter $\sigma_{_{\rm NR}}$ to be
 \begin{eqnarray} \label{eq:sigma}
  \sigma_{_{\rm NR}}= e^{i\pi/4}\left(3.36^{+1.12}_{-0.96}\right)\,{\rm GeV}.
 \end{eqnarray}
It turns out that the nonresonant contribution arises dominantly
from the transition process (88\%) via the scalar-density-induced
vacuum to $K\bar K$ transition, namely, $\langle K^+K^-|\bar
ss|0\rangle$, and slightly from the current-induced process (3\%).
Physically, this is because the decay $B\to KKK$ is dominated by the
$b\to s$ penguin transition. The nonresonant background in $B\to KK$
transition does not suffice to account for the experimental
observation that the penguin-dominated decay $B\to KKK$ is dominated
by the nonresonant contributions. This implies that the two-body
matrix element e.g. $\langle K\overline K|\bar ss|0\rangle$ induced
by the scalar density should have a large nonresonant component.

We have considered other $B\to KKK$ decays such as $B^-\to
K^+K^-K^-$ and $B^-\to K^-K_SK_S$ and found that they are also
dominated by the nonresonant contributions. Our predicted branching
ratio ${\cal B}(B^-\to K^+K^-K^-)_{\rm
NR}=(25.3^{+4.9}_{-4.5})\times 10^{-6}$ is in good agreement with
the Belle measurement of $(24.0^{+3.0}_{-6.2})\times 10^{-6}$, but a
factor of 2 smaller than the BaBar result of $(50\pm6\pm4)\times
10^{-6}$ (see Table 1).

The resonant and nonresonant contributions to the decay $B^-\to
K^-\pi^+\pi^-$ are shown in Table \ref{tab:Kpipi}. We see that the
calculated $K^*\pi$ and $\rho K$ rates are smaller than the data by
a factor of $2\sim 3$. This seems to be a generic feature of the
factorization approach such as QCD factorization where the predicted
penguin-dominated $VP$ rates are too small compared to experiment.
We shall return back to this point later.

While Belle has found a sizable fraction of order $(35\sim40)\%$ for
the nonresonant signal in $K^-\pi^+\pi^-$ and $\overline
K^0\pi^+\pi^-$ modes, BaBar reported a small fraction of order 4.5\%
in $K^-\pi^+\pi^-$ (see Table \ref{tab:BRexpt}). The huge disparity
between BaBar and Belle is ascribed to the different
parameterizations adopted by both groups. While Belle
\cite{BelleKpipi} employed the parametrization Eq. (\ref{eq:ANR}) to
describe the nonresonant contribution, BaBar \cite{BaBarKpipi} used
the LASS parametrization to describe the $K\pi$ $S$-wave and the
nonresonant component by a single amplitude suggested by the LASS
collaboration to describe the scalar amplitude in elastic $K\pi$
scattering.

\begin{table}[t]
\tbl{Branching ratios (in units of $10^{-6}$) of resonant and
nonresonant (NR) contributions to $B^-\to K^-\pi^+\pi^-$.
Theoretical errors correspond to the uncertainties in (i)
$\alpha_{_{\rm NR}}$, (ii) $m_s$, $F^{BK}_0$ and $\sigma_{_{\rm
NR}}$, and (iii) $\gamma=(59\pm7)^\circ$. We do not have $1/m_b$
power corrections within this model.}
%\begin{ruledtabular}
{\begin{tabular}{@{}l l l l@{}}  \toprule
 Decay mode~~& BaBar \cite{BaBarKpipi} & Belle \cite{BelleKpipi} &
 Theory \cite{CCS3body}
\\ \colrule
 $\overline K^{*0}\pi^-$ & $9.04\pm0.77\pm0.53^{+0.21}_{-0.37}$
 & $6.45\pm0.43\pm0.48^{+0.25}_{-0.35}$ &  $3.0^{+0.0+0.8+0.0}_{-0.0-0.7-0.0}$ \\
 $\overline K^{*0}_0(1430)\pi^-$ & $34.4\pm1.7\pm1.8^{+0.1}_{-1.4}$ &
$32.0\pm1.0\pm2.4^{+1.1}_{-1.9}$ & $10.5^{+0.0+3.2+0.0}_{-0.0-2.7-0.1}$  \\
 $\rho^0K^-$ & $5.08\pm0.78\pm0.39^{+0.22}_{-0.66}$ &
$3.89\pm0.47\pm0.29^{+0.32}_{-0.29}$ & $1.3^{+0.0+1.9+0.1}_{-0.0-0.7-0.1}$ \\
 $f_0(980)K^-$ & $9.30\pm0.98\pm0.51^{+0.27}_{-0.72}$
 & $8.78\pm0.82\pm0.65^{+0.55}_{-1.64}$ & $7.7^{+0.0+0.4+0.1}_{-0.0-0.8-0.1}$  \\
NR & $2.87\pm0.65\pm0.43^{+0.63}_{-0.25}$ &
$16.9\pm1.3\pm1.3^{+1.1}_{-0.9}$ & $18.7^{+0.5+11.0+0.2}_{-0.6-~6.3-0.2}$ \\
\colrule
 Total & $64.4\pm2.5\pm4.6$ & $48.8\pm1.1\pm3.6$ & $45.0^{+0.3+16.4+0.1}_{-0.4-10.5-0.1}$ \\
 \botrule
\end{tabular} \label{tab:Kpipi}}
%\end{ruledtabular}
\end{table}

From Table \ref{tab:Kpipi} we see that our predicted nonresonant
rates are in agreement with the Belle measurements but larger than
the BaBar result. The reason for the large nonresonant rates in the
$K^-\pi^+\pi^-$ mode is that under SU(3) flavor symmetry, we have
the relation $\langle K\pi|\bar sq|0\rangle^{NR}=\langle K\bar
K|\bar ss|0\rangle^{NR}$.
Hence, the nonresonant rates in the $K^-\pi^+\pi^-$ and $\overline
K^0\pi^+\pi^-$ modes should be similar to that in $K^+K^-\overline
K^0$ or $K^+K^-K^-$. Since the $KKK$ channel receives resonant
contributions only from $\phi$ and $f_{0}$ mesons, while $K^*,
K^*_{0},\rho,f_{0}$ resonances contribute to $K\pi\pi$ modes, this
explains why the nonresonant fraction is of order 90\% in the former
and becomes of order 40\% in the latter. It is interesting to notice
that, based on a simple fragmentation model and SU(3) symmetry,
Gronau and Rosner \cite{Gronau3body} also found  a large nonresonant
background in $K^-\pi^+\pi^-$ and $\overline K^0\pi^+\pi^-$.

Very recently, BaBar has reported the measurement of
the nonresonant contribution in the $K^-\pi^+\pi^0$ mode \cite{BaBarKppimpi0}. It is clear that our prediction is
larger than the BaBar result and barely consistent with the
Belle limit (see Table \ref{tab:Kmpippi0}). As stressed in \cite{CCS3body}, it is conceivable that the SU(3) breaking effect in
$\langle K\pi|\bar sq|0\rangle^{NR}$ may lead to a result consistent with both BaBar and Belle.

\begin{table}[h]
\tbl{Branching ratios (in units of $10^{-6}$) of resonant and
nonresonant (NR) contributions to $\overline B^0\to K^-\pi^+\pi^0$. In the BaBar measurement, the resonance $K_0^*(1430)$ is replaced by the $S$-wave $K\pi$ state, namely, $(K\pi)_0^*$.}
%\begin{ruledtabular} 
{\begin{tabular}{@{}l l l l@{}}  \toprule
 Decay mode~~ & BaBar \cite{BaBarKppimpi0} & Belle \cite{BelleKppimpi0} & Theory  \\ \colrule
 $K^{*-}\pi^+$ & $4.2^{+0.9}_{-0.5}\pm0.3$  & $4.9^{+1.5+0.5+0.8}_{-1.5-0.3-0.3}$ &   $1.0^{+0.0+0.3+0.1}_{-0.0-0.3-0.1}$ \\
 $\overline K^{*0}\pi^0$ & $2.4\pm0.5\pm0.3$  & $<2.3$ &  $1.0^{+0.0+0.3+0.2}_{-0.0-0.2-0.1}$  \\
 $K^{*-}_0(1430)\pi^+$ & $8.5^{+1.0+1.3}_{-1.2-1.0}\pm1.6$ &  $5.1\pm1.5^{+0.6}_{-0.7}$ & $5.0^{+0.0+1.5+0.1}_{-0.0-1.3-0.1}$ \\
 $\overline K^{*0}_0(1430)\pi^0$ & $7.8^{+1.0+1.6}_{-0.8-1.2}\pm2.0$ &  $6.1^{+1.6+0.5}_{-1.5-0.6}$ & $4.2^{+0.0+1.4+0.0}_{-0.0-1.2-0.0}$ \\
 $\rho^+K^-$ & $8.0^{+0.8}_{-1.3}\pm0.6$ & $15.1^{+3.4+1.4+2.0}_{-3.3-1.5-2.1}$ & $2.5^{+0.0+3.6+0.2}_{-0.0-1.4-0.2}$ \\
 NR & $4.4\pm0.9\pm0.5$ & $5.7^{+2.7+0.5}_{-2.5-0.4}<9.4$ & $9.6^{+0.3+6.6+0.0}_{-0.2-3.5-0.0}$ \\ \colrule
 Total & $35.7^{+2.6}_{-1.5}\pm2.2$ & $36.6^{+4.2}_{-4.1}\pm3.0$ & $28.9^{+0.2+16.1+0.2}_{-0.2-~9.4-0.2}$ \\
  \botrule
\end{tabular} \label{tab:Kmpippi0} }
%\end{ruledtabular}
\end{table}

\section{Tree-dominated $B\to \pi\pi\pi,K\pi\pi$ modes}
The $B\to \pi\pi\pi$ mode receives nonresonant contributions mostly
from the $b\to u$ transition as the nonresonant contribution in the
penguin matrix element $\langle\pi^+\pi^-|\bar dd|0\rangle$ is
suppressed by the smallness of penguin Wilson coefficients $a_6$ and
$a_8$. This indicates that the nonresonant fraction, of order 15\%
in the decay $B^-\to\pi^+\pi^-\pi^-$, is small in the tree-dominated
three-body $B$ decays.

Note that while $B^-\to \pi^+\pi^-\pi^-$ is dominated by the
$\rho^0$ pole, the decay $\overline B^0\to \pi^+\pi^-\pi^0$ receives
$\rho^\pm$ and $\rho^0$ contributions. As a consequence, the
$\pi^+\pi^-\pi^0$ mode has a rate larger than $\pi^+\pi^-\pi^-$ even
though the former involves a $\pi^0$ in the final state.

Among the 3-body decays we have studied, the decay $B^-\to
K^+K^-\pi^-$ dominated by $b\to u$ tree transition and $b\to d$
penguin transition has the smallest branching ratio of order
$4\times 10^{-6}$. BaBar \cite{BaBar:KKpi} has recently reported
the observation of the decay $B^+\to K^+K^-\pi^+$ with the
branching ratio $(5.0\pm0.5\pm0.5)\times 10^{-6}$. Our prediction
for this mode, $(4.0^{+0.5+0.7+0.3}_{-0.6-0.5-0.3})\times 10^{-6}$,
is in accordance with experiment.

\section{Quasi-two-body $B$ decays}
It is known that in the narrow width approximation, the 3-body
decay rate obeys the factorization relation
 \begin{eqnarray} \label{eq:fact}
 \Gamma(B\to RP\to P_1P_2P)=\Gamma(B\to RP){\cal B}(R\to P_1P_2),
 \end{eqnarray}
with $R$ being a vector meson or a scalar resonance. We have
computed the resonant contributions to 3-body decays and determined
the rates for the quasi-two-body decays $B\to VP$ and $B\to SP$. The
predicted $\rho\pi,~f_0(980)K$ and $f_0(980)\pi$ rates are in
agreement with the data, while the calculated $\phi K,~K^*\pi,~\rho
K$ and $K_0^*(1430)\pi$ are in general too small compared to
experiment. The fact that this work and QCDF lead to too small rates
for $\phi K,~K^*\pi,~\rho K$ and $K_0^*(1430)\pi$ may imply the
importance of power corrections due to the non-vanishing $\rho_A$
and $\rho_H$ parameters arising from weak annihilation and hard
spectator interactions, respectively, which are used to parametrize
the endpoint divergences, or due to possible final-state
rescattering effects from charm intermediate states \cite{CCSfsi}.
However, this is beyond the scope of the present work.

\begin{table}[ht]
\tbl{Branching ratios of quasi-two-body decays $B\to VP$ and $B\to
SP$ obtained from the studies of three-body decays based on the
factorization approach $^{4}$.  Theoretical uncertainties have been
added in quadrature. QCD factorization predictions taken from
$^{10}$ for $VP$ modes and from $^{11}$
 for $SP$ channels are shown here for comparison.}
%\begin{ruledtabular} \label{tab:BR2body}
{\begin{tabular}{l c c | c c} \toprule
 Decay mode~~ &  BaBar & Belle  & QCDF & This work  \\ \colrule
 $\phi K^0$ & $8.4^{+1.5}_{-1.3}\pm0.5$   &
 $9.0^{+2.2}_{-1.8}\pm0.7$   & $4.1^{+0.4+1.7+1.8+10.6}_{-0.4-1.6-1.9-~3.0}$ & $5.3^{+1.0}_{-0.9}$ \\
 $\phi K^-$ & $8.4\pm0.7\pm0.7$ & $9.60\pm0.92^{+1.05}_{-0.84}$ &
 $4.5^{+0.5+1.8+1.9+11.8}_{-0.4-1.7-2.1-~3.3}$ & $5.9^{+1.1}_{-1.0}$ \\
 $\overline K^{*0}\pi^-$ & $13.5\pm1.2^{+0.8}_{-0.9}$ &
 $9.8\pm0.9^{+1.1}_{-1.2}$ & $3.6^{+0.4+1.5+1.2+7.7}_{-0.3-1.4-1.2-2.3}$ & $4.4^{+1.1}_{-1.0}$ \\
 $\overline K^{*0}\pi^0$ & $3.0\pm0.9\pm0.5$ & $<3.5$ & $0.7^{+0.1+0.5+0.3+2.6}_{-0.1-0.4-0.3-0.5}$
 & $1.5^{+0.5}_{-0.4}$ \\
 $K^{*-}\pi^+$ & $11.0\pm1.5\pm0.7$ & $8.4\pm1.1^{+0.9}_{-0.8}$ &
 $3.3^{+1.4+1.3+0.8+6.2}_{-1.2-1.2-0.8-1.6}$ & $3.1^{+0.9}_{-0.9}$ \\
 $K^{*-}\pi^0$ & $6.9\pm2.0\pm1.3$  & & $3.3^{+1.1+1.0+0.6+4.4}_{-1.0-0.9-0.6-1.4}$ &
 $2.2^{+0.6}_{-0.5}$ \\
 $K^{*0}K^-$ & & &
 $0.30^{+0.11+0.12+0.09+0.57}_{-0.09-0.10-0.09-0.19}$ & $0.35^{+0.06}_{-0.06}$ \\
 $\rho^0K^-$ & $5.1\pm0.8^{+0.6}_{-0.9}$ &
 $3.89\pm0.47^{+0.43}_{-0.41}$ & $2.6^{+0.9+3.1+0.8+4.3}_{-0.9-1.4-0.6-1.2}$ & $1.3^{+1.9}_{-0.7}$  \\
 $\rho^0\overline K^0$ & $4.9\pm0.8\pm0.9$ & $6.1\pm1.0\pm1.1$
 & $4.6^{+0.5+4.0+0.7+6.1}_{-0.5-2.1-0.7-2.1}$ & $2.0^{+1.9}_{-0.9}$ \\
 $\rho^+K^-$ & $8.6\pm1.4\pm1.0$ & $15.1^{+3.4+2.4}_{-3.3-2.6}$ &
 $7.4^{+1.8+7.1+1.2+10.7}_{-1.9-3.6-1.1-~3.5}$  & $2.5^{+3.6}_{-1.4}$ \\
 $\rho^-\overline K^0$ & $8.0^{+1.4}_{-1.3}\pm0.5$  & &
 $5.8^{+0.6+7.0+1.5+10.3}_{-0.6-3.3-1.3-~3.2}$ & $1.3^{+3.0}_{-0.9}$ \\
 $\rho^0\pi^-$ & $8.8\pm1.0^{+0.6}_{-0.9}$ &
 $8.0^{+2.3}_{-2.0}\pm0.7$   & $11.9^{+6.3+3.6+2.5+1.3}_{-5.0-3.1-1.2-1.1}$  & $7.7^{+1.7}_{-1.6}$ \\
 $\rho^-\pi^+$ & & &
 $21.2^{+10.3+8.7+1.3+2.0}_{-~8.4-7.2-2.3-1.6}$ & $15.5^{+4.0}_{-3.5}$ \\
 $\rho^+\pi^-$ & & & $15.4^{+8.0+5.5+0.7+1.9}_{-6.4-4.7-1.3-1.3}$ & $8.5^{+1.1}_{-1.0}$  \\
 $\rho^0\pi^0$ & $1.4\pm0.6\pm0.3$ & $3.1^{+0.9+0.6}_{-0.8-0.8}$ &
 $0.4^{+0.2+0.2+0.9+0.5}_{-0.2-0.1-0.3-0.3}$ & $1.0^{+0.3}_{-0.2}$ \\
 $f_0(980)K^0;f_0\to \pi^+\pi^-$ & $5.5\pm0.7\pm0.6$ & $7.6\pm1.7^{+0.8}_{-0.9}$ &
 $6.7^{+0.1+2.1+2.3}_{-0.1-1.5-1.1}$   & $7.7^{+0.4}_{-0.7}$ \\
 $f_0(980)K^-;f_0\to \pi^+\pi^-$ & $9.3\pm1.0^{+0.6}_{-0.9}$ &
 $8.8\pm0.8^{+0.9}_{-1.8}$ & $7.8^{+0.2+2.3+2.7}_{-0.2-1.6-1.2}$  & $7.7^{+0.4}_{-0.8}$ \\
 $f_0(980)K^0;f_0\to K^+K^-$ & $5.3\pm2.2$  &  & & $5.8^{+0.1}_{-0.5}$ \\
 $f_0(980)K^-;f_0\to K^+K^-$ & $6.5\pm2.5\pm1.6$ & $<2.9$ &  & $7.0^{+0.4}_{-0.7}$ \\
 $f_0(980)\pi^-;f_0\to \pi^+\pi^-$ & $<3.0$ & & $0.5^{+0.0+0.2+0.1}_{-0.0-0.1-0.0}$
 & $0.39^{+0.03}_{-0.02}$ \\
 $f_0(980)\pi^-;f_0\to K^+K^-$ & & & & $0.50^{+0.06}_{-0.04}$ \\
 $f_0(980)\pi^0;f_0\to \pi^+\pi^-$ & & &
 $0.02^{+0.01+0.02+0.04}_{-0.01-0.00-0.01}$  & $0.010^{+0.003}_{-0.002}$ \\
 $\overline K^{*0}_0(1430)\pi^-$ & $36.6\pm1.8\pm4.7$ &
 $51.6\pm1.7^{+7.0}_{-7.4}$ & $11.0^{+10.3+7.5+49.9}_{-~6.0-3.5-10.1}$ & $16.9^{+5.2}_{-4.4}$ \\
 $\overline K^{*0}_0(1430)\pi^0$ & $12.7\pm2.4\pm4.4$ & $9.8\pm2.5\pm0.9$ &
 $6.4^{+5.4+2.2+26.1}_{-3.3-2.1-~5.7}$ & $6.8^{+2.3}_{-1.9}$ \\
 $K^{*-}_0(1430)\pi^+$ & $36.1\pm4.8\pm11.3$ & $49.7\pm3.8^{+4.0}_{-6.1}$
 & $11.3^{+9.4+3.7+45.8}_{-5.8-3.7-~9.9}$  & $16.2^{+4.7}_{-4.0}$ \\
 $K^{*-}_0(1430)\pi^0$ & & & $5.3^{+4.7+1.6+22.3}_{-2.8-1.7-~4.7}$
 &  $8.9^{+2.6}_{-2.2}$ \\
 $K^{*0}_0(1430)K^-$ & $<2.2$  & & & $1.3^{+0.3}_{-0.3}$ \\
 \botrule
\end{tabular} }
%\end{ruledtabular}
\end{table}

\section{Time-dependent $CP$ asymmetries}
The penguin-induced three-body decays $B^0\to K^+K^-K_S$ and
$K_SK_SK_S$ deserve special attention as the current measurements of
the deviation of $\sin 2\beta_{\rm eff}$ in $KKK$ modes from $\sin 2
\beta_{J/\psi K_S}$ may indicate New Physics in $b\to s$
penguin-induced modes.  It is of great importance to examine and
estimate how much of the deviation of $\sin 2\beta_{\rm eff}$ is
allowed in the SM. Owing to the presence of color-allowed tree
contributions in $B^0\to K^+K^-K_{S}$, this mode is subject to a
potentially significant tree pollution and the deviation of the
mixing-induced $CP$ asymmetry from that measured in $B\to J/\psi
K_S$ could be as large as ${\cal O}(0.10)$. Since the tree amplitude
is tied to the nonresonant background, it is very important to
understand the nonresonant contributions in order to have a reliable
estimate of $\sin 2\beta_{\rm eff}$ in $KKK$ modes.

\begin{table}[t]
\tbl{Mixing-induced and direct $CP$ asymmetries for various
charmless 3-body $B$ decays. Experimental results are taken from
$^{1}$.}
%\begin{ruledtabular}
{\begin{tabular}{@{}l r r r | r r @{} }  \toprule
 Decay & $\sin 2\beta_{\rm eff}$ & $\Delta\sin 2\beta_{\rm eff}$ & Expt &
 $A_f(\%)$ & Expt \\ \colrule
$K^+K^-K_S$ & $0.728^{+0.001+0.002+0.009}_{-0.002-0.001-0.020}$ &
$0.041^{+0.028}_{-0.033}$ & $0.05\pm0.11$ & $-4.63^{+1.35+0.53+0.40}_{-1.01-0.54-0.34}$ & $-7\pm8$  \\
$K_SK_SK_S$ & $0.719^{+0.000+0.000+0.008}_{-0.000-0.000-0.019}$ &
$0.039^{+0.027}_{-0.032}$ & $-0.10\pm0.20$ & $0.69^{+0.01+0.01+0.05}_{-0.01-0.03-0.07}$ & $14\pm15$ \\
$K_S\pi^0\pi^0$ & $0.729^{+0.000+0.001+0.009}_{-0.000-0.001-0.020}$
& $0.049^{+0.027}_{-0.032}$ & $-1.20\pm0.41$ & $0.28^{+0.09+0.07+0.02}_{-0.06-0.06-0.02}$ & $-18\pm22$ \\
$K_S\pi^+\pi^-$ & $0.718^{+0.001+0.017+0.008}_{-0.001-0.007-0.018}$
& $0.038^{+0.031}_{-0.032}$ & & $4.94^{+0.03+0.03+0.32}_{-0.02-0.05-0.40}$ & \\
 \botrule
\end{tabular} \label{tab:CP}}
%\end{ruledtabular}
\end{table}

The deviation of the mixing-induced $CP$ asymmetry in $B^0\to
K^+K^-K_S$, $K_SK_SK_S$, $K_S\pi^+\pi^-$ and $K_S\pi^0\pi^0$ from
that measured in $B\to \phi_{c\bar c}K_S$, i.e. $\sin 2
\beta_{\phi_{c\bar c}K_S}=0.681\pm0.025$ \cite{HFAG}, namely,
$\Delta \sin 2\beta_{\rm eff}\equiv \sin 2\beta_{\rm eff}-\sin 2
\beta_{\phi_{c\bar c}K_S}$, is summarized in Table \ref{tab:CP}. Our
calculation indicates the deviation of the mixing-induced $CP$
asymmetry in $\overline B^0\to K^+K^-K_{S}$ from that measured in
$\overline B^0\to \phi_{c\bar c}K_S$ is very similar to that of the
$K_SK_SK_S$ mode as the tree pollution effect in the former is
somewhat washed out. Nevertheless, direct $CP$ asymmetry of the
former, being of order $-4\%$, is more prominent than the latter.

\section*{Acknowledgments}

I'm grateful to Chun-Khiang Chua and Amarjit Soni for fruitful
collaboration and to the organizer Chun Liu for organizing this
stimulating conference.

%\begin{thebibliography}{000} %for 3 digits
%\begin{thebibliography}{00}  %for 2 digits


\begin{thebibliography}{00}
%\newcommand{\bi}{\bibitem}

\bibitem{HFAG} Heavy Flavor Averaging Group,
 http://www.slac.stanford.edu/xorg/hfag.

\bibitem{PDG} Particle Data Group, Y.M. Yao {\it et al.,} J. Phys. G
{\bf 33}, 1 (2006).

\bibitem{Fajfer} S. Fajfer, R.J. Oakes, and T.N. Pham, Phys. Rev. D {\bf 60}, 054029 (1999);  Phys.Lett. B {\bf 539}, 67 (2002);
S. Fajfer, T.N. Pham, and A. Prapotnik, Phys. Rev. D {\bf
70}, 034033 (2004).

\bibitem{Cheng:2002qu}
  H.Y.~Cheng and K.C.~Yang,
  %``Nonresonant three-body decays of D and B mesons,''
  Phys.\ Rev.\ D {\bf 66}, 054015 (2002).
  %%CITATION = HEP-PH 0205133;%%

\bibitem{CCS3body} H.Y. Cheng, C.K. Chua, and A. Soni,
Phys. Rev. D {\bf 76}, 094006 (2007).

\bibitem{CCSKKK} H.Y. Cheng, C.K. Chua, and A. Soni, Phys. Rev. D {\bf
72}, 094003 (2005).

\bibitem{BaBarKpipi} B. Aubert {\it et al.} (BaBar Collaboration), Phys.
Rev. D {\bf 72}, 072003 (2005).

\bibitem{BelleKpipi} A. Garmash {\it et al.} (Belle Collaboration), Phys.
Rev. Lett. {\bf 96}, 251803 (2006); Phys. Rev. D {\bf 75}, 012006
(2007).

\bibitem{Gronau3body} M. Gronau and J.L. Rosner, Phys. Rev. D {\bf 72}, 094031 (2005).

\bibitem{BaBarKppimpi0} B. Aubert {\it et al.} (BaBar Collaboration),
arXiv:0711.4417 [hep-ex].

\bibitem{BelleKppimpi0} P. Chang {\it et al.} (Belle Collaboration),
Phys. Lett. B {\bf 599}, 148 (2004).


\bibitem{BaBar:KKpi} B. Aubert {\it et al.}  (BaBar Collaboration), Phys. Rev. Lett. {\bf 99}, 221801 (2007).

\bibitem{BN} M. Beneke and M. Neubert, Nucl. Phys. B {\bf 675}, 333 (2003).


\bibitem{CCY} H.Y. Cheng, C.K. Chua, and K.C. Yang, Phys. Rev. D {\bf 73},
014017 (2006).


\bibitem{CCSfsi} H.Y.~Cheng, C.K.~Chua, and A.~Soni, Phys. Rev. D {\bf
71}, 014030 (2005).

\end{thebibliography}
\end{document}